\documentclass[a4paper,11pt]{article}
\pdfoutput=1  
\usepackage{jcappub}
\usepackage{graphicx,amssymb,amsmath,color}
\usepackage{hyperref}
\usepackage{multirow}

\newcommand{\beq}{\begin{equation}}
\newcommand{\eeq}{\end{equation}}
\def\bea{\begin{eqnarray}}
\def\eea{\end{eqnarray}}
\newcommand{\bei}{\begin{itemize}}
\newcommand{\eei}{\end{itemize}}
\def\bmat{\begin{matrix}}
\def\emat{\end{matrix}}

\newcommand{\Fig}[1]{Fig.~\ref{#1}}
\newcommand{\Eq}[1]{Eq.~(\ref{#1})}
\newcommand{\Eqs}[2]{Eqs.~(\ref{#1}) and (\ref{#2})}
\newcommand{\Sec}[1]{Sec.~\ref{#1}}

\newcommand{\nn}{\nonumber\\}
\newcommand{\aln}[1]{\begin{align}#1\end{align}}

\def\={\,=\,}
\def\+{\,+\,}
\def\-{\,-\,}

\def\Mpl{M_{\rm Pl}}

\begin{document}

\title{Low-energy probes of small CMB amplitude in models of radiative Higgs mechanism}

\author[a,b]{Sunghoon Jung}
\emailAdd{sunghoonj@snu.ac.kr}
\affiliation[a]{Center for Theoretical Physics, Department of Physics and Astronomy, Seoul National University, Seoul 08826, Korea}
\affiliation[b]{Astronomy Research Center, Seoul National University, Seoul 08826, Korea}

\author[a]{Kiyoharu Kawana}
\emailAdd{kawana@snu.ac.kr}

\abstract{
The small CMB amplitude $A_s \simeq 10^{-9}$ (or, small temperature fluctuation $\delta T/T \simeq 10^{-5}$) typically requires an unnaturally small effective coupling of an inflaton $\lambda_\phi \sim 10^{-14}$. In successful models, there usually is extra suppression of the amplitude, e.g. by large-field inflaton with non-minimal coupling $\xi$, so that $\lambda_\phi$ can be much larger. But $\lambda_\phi$ and $\xi$ cannot be $\sim {\cal O}(1)$ simultaneously; the naturalness burden is shared between them.
We show that the absence of new physics signals at TeV scale may prefer a more natural size of $\xi \lesssim {\cal O}(1-100)$ with $\lambda_\phi \lesssim {\cal O}(10^{-4}-10^{-8})$, constraining larger $\xi$ with larger $\lambda_\phi$ more strongly. 
This intriguing connection between low- and high-energy physics is made in the scenarios with $U(1)_X$ where inflaton's renormalization running also induces Coleman-Weinberg mechanism for the electroweak symmetry breaking. We particularly work out the prospects of LHC 13 and 100 TeV $pp$ colliders for probing the parameter space of the small CMB amplitude.
}

\maketitle

\section{Introduction}

Although slow-roll is almost an inevitable requirement of inflation, the observed small temperature fluctuation in the universe (or, more precisely the small CMB amplitude) is not. Slow-roll inflation could have explained homogeneity, causality, and the quantum origin of density perturbations without predicting the small CMB amplitude. The observed $\delta T/T \sim 10^{-5}$ (or $A_s \sim 10^{-9}$)~\cite{Aghanim:2018eyx,Akrami:2018odb} is typically translated to $\lambda_\phi \sim 10^{-14} \ll 1$ for an effective quartic potential description $V = \lambda_\phi \phi^4/4$ of an inflaton $\phi$. Such a small coupling is seemingly unnatural even though it is not a necessary prediction of slow-roll. Why is our universe realized so? Could this small CMB amplitude be related to other physics?

Anthropically, it was argued that if $\delta T/T$ were larger, there would have been too much black hole formations; while if smaller, there would have been too little structures within the age of the universe~\cite{Tegmark:1997in}. It was even argued that smaller $\delta T/T$ is more likely realized since it usually requires a flatter inflaton potential, inducing longer inflation and exponentially more numbers of Hubble patches~\cite{Vilenkin:1994ua}.

Higgs inflation has shed some light in this regard. The measured Higgs boson mass (combined with the top quark mass) surprisingly implies an almost vanishing Higgs potential near the Planck scale~\cite{Buttazzo:2013uya,EliasMiro:2011aa,Degrassi:2012ry,Bednyakov:2015sca,Hamada:2012bp}, via the renormalization group (RG) evolution in the Standard Model (SM). This not only made the electroweak vacuum possibly being metastable, but also made the case of the Higgs boson as an inflaton~\cite{Bezrukov:2007ep,Bezrukov:2009db,Bezrukov:2010jz,Salvio:2013rja} more plausible. As the first derivative of the Higgs potential is also expected to vanish near the same scale, some of the slow-roll conditions as well as small potential height are automatically satisfied.
Although Higgs inflation in its minimal form in the SM is not completely successful (so that generalized with an approximate inflection point~\cite{Hamada:2015wea,Ballesteros:2015noa,Choi:2016eif,Okada:2016ssd,Okada:2017cvy}, or a non-minimal coupling $\xi$~\cite{Hamada:2014wna,Barvinsky:2008ia,DeSimone:2008ei,Allison:2013uaa,Hamada:2013mya,Hamada:2014iga,Bezrukov:2014bra,Hamada:2014xka,Hamada:2017sga,Hamada:2021jls}, or various other features~\cite{Park:2008hz,Ferrara:2013rsa,Kallosh:2013yoa,Galante:2014ifa,Kallosh:2013hoa}), it became noticed that small CMB amplitude may have some connections with other physics at distant energy scales.

Further, this interesting coincidence between high- and low-energy potential shapes was attempted to be explained by multiple point principle~\cite{Froggatt:1995rt,Froggatt:2001pa,Nielsen:2012pu,Kawai:2011rj,Kawai:2011qb,Kawai:2013wwa,Hamada:2014ofa,Hamada:2014xra,Hamada:2015dja,Hamada:2015ria,Kannike:2020qtw} (i.e., why we seem to have almost degenerate minima at the electroweak and Planck scales) or by classical conformal invariance at the Planck scale (i.e., why the dimensionful Higgs quadratic term almost vanishes there). These hypotheses triggered a possibility of all dimensionful parameters in the SM being induced by RG evolutions. In particular, electroweak symmetry breaking (EWSB) was successfully induced starting from even a vanishing Higgs potential (all potential terms vanish at the Planck scale) just with an additional $U(1)$ gauge symmetry, a charged fermion, and a symmetry breaking scalar~\cite{Chun:2013soa,Hashimoto:2013hta}. Also, it was shown that RG evolution can be used to realize (saddle-point) inflation together with the spontaneous symmetry breaking of $U(1)_{B-L}$ at a low scale~\cite{Kawana:2015tka,Choi:2016eif,Okada:2016ssd,Okada:2017cvy}. 

In this paper, along this light, we study whether small CMB amplitude can have meaningful connections or correlations with low-energy physics in the models where $\phi$ inflaton's renormalization running induces both slow-roll inflation and radiative breaking of the electroweak symmetry (Coleman-Weinberg mechanism). These well measured high- and low-energy physics may induce non-trivial constraints on $\phi$ through perturbative quantum corrections (RG evolutions). In \Sec{sec:CW}, we start by introducing our model and describing CW mechanism, then in \Sec{sec:inflationmodel} we match this model to simple inflation models, and in \Sec{sec:results} we interpret low-energy searches to the constraints on the parameter space of inflation sector that can explain the small CMB amplitude.

\section{Coleman-Weinberg mechanism with $U(1)_X$} \label{sec:CW}

Our model consists of Coleman-Weinberg (CW) mechanism for the electroweak symmetry breaking (EWSB) with vanishing dimensionful parameters at $\Mpl$. This may represent one class of resolutions to the Planck-weak hierarchy problem. 

CW mechanism is realized with an additional $U(1)_X$,  a linear combination of $U(1)_{B-L}$ and $U(1)_Y$ hypercharge, parameterized by $x$ as 
\beq
X \= (B-L) - x Y.
\eeq
This model in our framework has been worked out in \cite{Chun:2013soa} (including RG equations), with two small but necessary changes discussed below. 
To summarize, one generation (for simplicity) of right-handed neutrino $\nu_R$ and the SM-singlet scalar $\Phi$ with $X(\Phi)=2$ are introduced, having Yukawa interactions
\beq
{\cal L} \, \ni\, -y_\nu \bar{\ell} \nu_R H \- y_N \overline{(\nu_R)^c} \nu_R \Phi
\eeq
and scalar potentials and the portal mixing
\beq
V \, = \,  m_H^2 |H|^2 \+ m_\Phi^2 |\Phi|^2  \+ \lambda_h |H|^4 \+ \lambda_\phi |\Phi|^4 \+ \lambda_{h\phi} |H|^2 |\Phi|^2.
\eeq
In this work, quadratic terms vanish at $\Mpl$: $m_H^2(\Mpl)=0$ and $m_\Phi^2(\Mpl)=0$, and they are not even RG induced at lower scales (since we work with a mass-independent scheme of dimensional regularization) but only by spontaneous symmetry breaking as will be discussed. We use the unitary gauge with $H = h/\sqrt{2}$ and $\Phi = \phi/\sqrt{2}$.

The one-loop CW effective potential of $\phi$ is 
\bea
V(\phi) 
 &\=& \frac{1}{4} \lambda_\phi(\Mpl) \phi^4 + \frac{\phi^4}{64\pi^2} \left( 10 \lambda_\phi^2 + 48 g_X^4 -8 y_{N}^4 \right)  \ln \frac{\phi^2}{\Mpl^2}  \+ \Delta V.
\label{eq:CWpot} \eea
This can also be approximated in terms of the running coupling $\lambda_\phi(\phi)$ as
\beq
V(\phi) \, \simeq \, \frac{1}{4} (\lambda_\phi(\phi) +C) \phi^4 + \Delta V,
\label{eq:CWpot0} \eeq
with the running
\beq
\lambda_\phi(\phi) \,\simeq\, \lambda_\phi(\Mpl) + \beta_{\lambda_\phi} \ln \frac{\phi}{\Mpl}.
\label{eq:lambdarun} \eeq
We use $\overline{\rm MS}$ scheme $C=0$ so that $\lambda_\phi(\Mpl)$ directly measures the CMB amplitude $A_s \propto \lambda_\phi(\Mpl)$\footnote{In \cite{Chun:2013soa} focusing on the CW mechanism for the EWSB, $C=-\frac{25}{12} \beta_{\lambda_\phi}(v_\phi)$ was chosen by the renormalization condition at low-energy $\left. \frac{\partial^4 V}{\partial \phi^4} \right|_{\phi=v_\phi} = 6\lambda_\phi(v_\phi)$. $C$ effectively only shifts the value of $\lambda_\phi$.}, and $\Delta V$ will be determined soon. $\Mpl$ is assumed to be the high-energy inflation scale, for simplicity.
The form in \Eq{eq:CWpot0} is convenient in matching the potential at the inflation scale. It is a good approximation to the correct \Eq{eq:CWpot} as field-strength renormalizations (not explicitly appearing in \Eq{eq:CWpot}), denoted by $\cdots$ in
\beq
(4\pi)^2 \beta_{\lambda_\phi} \= 20 \lambda_\phi^2 + 96 g_X^4 -16 y_N^4  \+  \cdots,
\eeq
are small ${\cal O}(\lambda_\phi g_X^2, \lambda_\phi y_N^2)$.

The potential is minimized at $v_\phi$ ($U(1)_X$ spontaneously broken) when
\beq
\lambda_\phi (v_\phi) \=  -\left( \frac{\beta_{\lambda_\phi}}{4} + C \right) \= - \frac{1}{64\pi^2} \left( 20 \lambda_\phi^2 + 96 g_X^4 - 16 y_N^4 \right) (v_\phi),
\label{eq:minimization} \eeq 
where $C=0$.
This requires $\lambda_\phi(v_\phi) \sim g_X^4, y_N^4 \ll g_X^2, y_N^2$, and such a hierarchical structure can be RG induced. Consequently, $v_\phi$ is exponentially suppressed compared to $\Mpl$, by dimensional transmutation
\beq
v_\phi  \,\simeq\, \Mpl \, e^{-\frac{1}{4}} \exp\left( -\frac{4\pi^2 \lambda_\phi(\Mpl) }{ 24 g_X^4 (\Mpl) - 4y_N^4(\Mpl) } \right),  \label{eq:vphi}
\eeq
resolving the hierarchy problem -- it is the CW mechanism.
The symmetry breaking is subsequently transmitted to the SM Higgs potential through the portal mixing as
\beq
m_H^2 (v_\phi) \= \frac{1}{2}\lambda_{h\phi}(v_\phi) v_\phi^2.
\label{eq:mh2} \eeq
Thus, the negative $\lambda_{h\phi}(v_\phi)$ is needed to break the electroweak symmetry too; such a negative value will also be RG induced (\Eq{eq:betaportal}). The electroweak vev $v_{EW} \equiv \langle h \rangle$ is finally defined as
\beq
v_{\rm EW} \= \sqrt{- \frac{m_H^2(v_{\rm EW})}{\lambda_h(v_{\rm EW})}},
\label{eq:vew} \eeq
which is required to be 246 GeV.

The potential at the minimum $\phi=v_\phi$ is
\beq
V(\phi = v_\phi) \= \frac{v_\phi^4}{4}( \lambda_\phi(v_\phi) + C)  + \Delta V \= - \frac{v_\phi^4}{16} \beta_{\lambda_\phi}(v_\phi)  + \Delta V,
\label{eq:Vvphi} \eeq
where the second equality used the minimization condition \Eq{eq:minimization} and holds independently on $C$.
The potential is non-zero and negative ($\beta_{\lambda_\phi}(v_\phi) >0$) without $\Delta V$.  Thus, we set $\Delta V = + \frac{v_\phi^4}{16} \beta_{\lambda_\phi}(v_\phi)$ to make it zero and avoid unnecessary dark energy. This also necessarily affects the potential height at the inflation scale, but as will be discussed in the next section this is negligible at the inflation scale.

\medskip
By this mechanism, even a strict flatland scenario starting with  $\lambda_\phi(\Mpl)=0$ and $\lambda_{h\phi}(\Mpl)=0$ can successfully induce EWSB~\cite{Chun:2013soa,Hashimoto:2013hta}. $\lambda_\phi$ has to be positive before it breaks $U(1)_X$ so that $\beta_{\lambda_\phi}<0$ at high scales, but $\lambda_\phi$ must turn to be negative to break $U(1)_X$ requiring $\beta_{\lambda_\phi}$ to flip its sign at some intermediate scale. This flip is achieved by the fine interplay of bosonic $g_X$  and fermionic $y_N$ contributions to the beta function
\beq
(4\pi)^2 \beta_{\lambda_\phi} \,\simeq \, 96 g_X^4 -16 y_N^4 
\label{eq:betalambda} \eeq 
with their respective running 
\beq
(4\pi)^2 \beta_{g_X} \, \simeq \, \left( 12 - \frac{32}{3}x +\frac{41}{6}x^2 \right) g_X^3, \qquad (4\pi)^2\beta_{y_N} \,\simeq\, 6y_N(y_N^2- g_X^2).
\eeq
As a result, for almost any $g_X(\Mpl)$, there exists a unique solution $y_N(\Mpl)$ for successful EWSB. The resulting collider phenomena depend on the value of $g_X(\Mpl)$. As shown in \cite{Chun:2013soa}, the smaller $g_X(\Mpl)$ is, the smaller corresponding $y_N(\Mpl)$ is. And more relevantly,  $M_X = 2 g_X(v_\phi) v_\phi$ becomes larger and the Higgs mixing angle smaller (among many observables). But  there is no definite range of such predictions since any value of $g_X(\Mpl)$ can induce EWSB. This  is one crucial difference of our work.

In this paper, above all, we relax the condition $\lambda_\phi(\Mpl)=0$ in order to also explain inflation with the inflaton $\phi$. As will be discussed, this brings definite ranges of low-energy predictions, leading to intimate connection between high- and low-energy physics. \cite{Kawana:2015tka} considered such a scenario but with a different focus; \cite{Okada:2016ssd,Okada:2017cvy} without EWSB. Later, we also relax $\lambda_{h\phi}(\Mpl)=0$, mainly to explore the dependence of our conclusion on this model parameter. A wider range of $x$ is allowed now (see~\cite{Hashimoto:2013hta} for the allowed narrow range in a flatland), but only $x \sim 1$ will be considered to avoid too large $X$ charges; models with $x=4/5$ and $2$ can be obtained from the $SO(10)$ grand unified gauge group.

\section{Inflation models for interpretation}  \label{sec:inflationmodel}

In this section, we introduce benchmark models of inflation, which are simple enough to represent a large range of models and allow simple interpretations of low-energy results.
The inflaton potential must also match with the CW potential in \Eq{eq:CWpot} at low energy.

We start with a quartic chaotic inflation (as a warm up and to show basic features)
\aln{V(\phi) \= \frac{\lambda_\phi^{}(\mu=\phi)}{4}\phi^4 +\Delta V~,
\label{inflaton potential}
} 
where $\lambda_\phi^{}(\mu=\phi)$ is the running quartic coupling evaluated at $\mu=\phi$.
This potential obviously matches with the CW potential \Eq{eq:CWpot0} at low energy, again up to small field-strength renormalizations.
We first show that the effect from $\Delta V$ (introduced in \Eq{eq:CWpot0} and determined in \Eq{eq:Vvphi}) is negligible. By rewriting the potential as $V + \Delta V = \frac{\lambda_\phi}{4}(1 + \delta) \Mpl^4$, the fractional correction 
\beq
\delta \, \equiv \, \frac{\Delta V}{V (\Mpl)} \= \frac{\beta_{\lambda_\phi}(v_\phi)}{4 \lambda_\phi(\phi_I)} \left( \frac{v_\phi}{\Mpl} \right)^4 \, \ll \, 1
\eeq
is negligible for only mildly suppressed $v_\phi \lesssim \Mpl$ because $\beta_{\lambda_\phi} \sim g_X^4 \sim y_N^4$ and $\lambda_\phi(\Mpl) = 10^{-4,-8},\, g_X = 10^{-3} \sim 10^{-1}$ are relevant in this work (see next section).

This simple model shows that the CMB amplitude directly measures the potential height at the inflation scale
\beq
\quad A_s \= \frac{1}{24\pi^2}\frac{V}{\Mpl^4} \frac{1}{\epsilon} \,\simeq\, 5.76 \times 10^5 \,\frac{\lambda_\phi(\Mpl)}{\pi^2},
\eeq
so that very small $\lambda \simeq 4 \times 10^{-14}$ is needed to explain the observed Planck 2018 data~\cite{Aghanim:2018eyx,Akrami:2018odb} as alluded:
\aln{
A_s
	&\=	2.101^{+0.031}_{-0.034}\times 10^{-9}&
	&&
&(68\%\ {\rm CL}),
\label{eq:Asdata}}
where CMB observables are evaluated at the pivot scale $k_*^{}=0.05\,{\rm Mpc}^{-1}$ with $N= 60$.
But this minimal model predicts too large tensor-to-scalar ratio $r$ 
\aln{
r \= 16\varepsilon \,\simeq\, \frac{16}{N}~,\quad n_s \= 1-6\varepsilon+2\eta \,\simeq\, 1-\frac{3}{N}~, 
\label{eq:quartic}}
compared to the Planck data
\aln{
r	&\,<\,	0.056,&
&&
&(95\%\ {\rm CL})\nn
n_s 
	&\=	0.9665\pm 0.0038,&
\frac{dn_s }{d\ln k}
	&\=	0.013\pm 0.024&
&(68\%\ {\rm CL}).
	\label{values of cosmological observables}
}
%


\

\noindent $\bullet$ {\bf Inflation with non-minimal coupling} \\
Our main benchmark model is the quartic potential with a non-minimal coupling $\xi \phi^2 R/M_{pl}^2$ (with dimensionless $\xi >0$) between inflaton and gravity \cite{Bezrukov:2007ep,Hamada:2014wna,Barvinsky:2008ia,DeSimone:2008ei,Allison:2013uaa,Hamada:2013mya,Hamada:2014iga,Bezrukov:2014bra,Hamada:2014xka,Hamada:2017sga,Hamada:2021jls}. This model is known to realize successful inflation, and the single new parameter $\xi$ allows easy interpretation of our results.

This model works as $\xi$ effectively suppresses the quartic potential at the inflation scale 
\beq
V_E \= \frac{\lambda_\phi}{4} \frac{\phi^4}{\Omega^4}, \qquad \Omega^2 \=1 + \xi \left( \frac{\phi}{\Mpl} \right)^2 >1,
\eeq
where the subscript $E$ refers to the Einstein frame; a canonical normalization of $\phi$ brings additional modifications but this is a basic structure. Thus, $\xi$ becomes effective for large-field inflation $\phi \gtrsim \Mpl / \sqrt{\xi}$.
In this limit, (similarly to the conventional Higgs inflation case) CMB observables are 
\aln{
r \,\simeq \, \frac{12}{N^2}~,\quad n_s^{} \,\simeq\, 1-\frac{2}{N}~,\quad A_s^{}\,\simeq\, \frac{\lambda_\phi^{}(\Mpl)N^2}{72\pi^2 \xi^2}~. 
}
Not only can $r$ and $n_s$ be consistent with Planck data, but we also obtain the following relation between $\lambda_\phi^{}(M_{pl}^{})$ and $\xi$ by the CMB amplitude $A_s^{}\simeq 2.1\times 10^{-9}$ in \Eq{eq:Asdata}
\aln{
\frac{\lambda_\phi(\Mpl)}{\xi^2}\, \simeq\, 4.1 \times 10^{-10} \left( \frac{60}{N} \right)^2.
\label{eq:xi}
}
The larger $\lambda_\phi(\Mpl)$, the larger field value at the inflation scale, hence the larger suppression by $\xi$ is needed. 

How large or small values of $\xi$ are natural, or most preferred? Since we do not specify a fundamental theory that might be able to calculate the value of $\xi$, it is naively reasonable to consider $\xi \sim {\cal O}(1)$ as a natural value. If we restrict $\xi \lesssim 100$ for example, the required size of $\lambda_\phi(\Mpl) \lesssim 10^{-6}$ can be significantly larger (hence, more natural or likely) than the naive translation $10^{-14}$ mentioned in the introduction, albeit still too small to be perfectly natural. 
On the other hand, there exists a lower bound on $\xi$. In the limit of $\xi \to 0$, the theory asymptotes to a pure quartic potential, which is inconsistent with observations as discussed in \Eq{eq:quartic}. We have numerically checked that $\xi \gtrsim 0.01$ in order to be consistent with CMB observations; the above large-$\phi$ approximation starts to break down for $\xi \lesssim 0.1$ (or, $\lambda_\phi(\Mpl) \lesssim 10^{-11}$). It is also possible to have $\lambda_\phi \sim {\cal O}(1)$ natural while with much larger $\xi \sim 10^5$, but usually too strong interactions can produce various unexpected corrections too. Thus, we will regard $0.01 \lesssim \xi \lesssim 100$ (or, $10^{-12} \lesssim \lambda_\phi(\Mpl) \lesssim 10^{-6}$) as the most desired (natural and comfortable) parameter space of inflation. Later, we will see that this is exactly the parameter space that is preferred by low-energy constraints.

\medskip
Lastly, we comment on the reheating after inflation. 
In usual $U(1)_X^{}$ models, the transfer of energy from inflatons to radiation can occur through perturbative decays $\phi\rightarrow Z'Z'~(\nu_R^{}\nu_R^{})$, or non-perturbative particle productions caused by the oscillation of $\phi$.  
For $\lambda_{\phi}^{}(\Mpl)\gtrsim 10^{-3}$ (or, $\xi\gtrsim 10^3)$, the qualitative behaviors of the preheating are expected to resemble those of Higgs inflation because $\phi$ couples to $Z'~(\nu_R^{})$ in the similar way as $H$ couples to $W,Z~(\nu_R^{})$.      
Thus, the reheating can occur instantaneously and the reheat temperature $T_R^{}$ can be high ${\cal O}(\rho_{\rm inf}^{1/4})={\cal O}(\lambda_\phi^{1/2}\Mpl /\xi^{1/2})$ \cite{DeCross:2015uza,DeCross:2016fdz,DeCross:2016cbs,Sfakianakis:2018lzf}. 
On the other hand, for $\lambda_{\phi}^{}(\Mpl)\ll 10^{-3}$ (or, $\xi\ll 10^3)$, the $U(1)_X^{}$ gauge coupling $g_X^{}$ can become much smaller than SM gauge couplings.
Such a small coupling prevents rapid perturbative decays of $Z'$ into SM fermions as well as reduces the efficiency of parametric resonances.  
But some of the conventional studies \cite{GarciaBellido:2008ab,Bezrukov:2008ut} still predict sufficiently large reheat temperature\footnote{For example, \cite{GarciaBellido:2008ab} shows that for $\xi\sim 1 (\lambda_\phi^{}\sim 10^{-8})$ and $g_X^{}\sim 0.01$, the ratio of energy densities between $Z'$ and $\phi$ is roughly given by $\sim 10^{-4}(2.7)^j/\sqrt{j}$, where $j$ is the number of zero crossings of $\phi$. This becomes ${\cal O}(1)$ for $j\sim 10$, which corresponds to large radiation energy $\rho_R^{}\sim \rho_\phi^{}\sim \lambda_\phi^{}\Mpl^4/(6\pi^2\xi^2 j^2)\sim (10^{-3}\Mpl)^4$.}, and \cite{Hamada:2020kuy} shows that the existence of higher dimensional operators such as $(\partial \phi)^4/(\Mpl/\sqrt{\xi})^2$ can significantly alter preheating dynamics. Thus, (p)reheating with non-minimal couplings is model dependent, and more dedicated estimations are beyond the scope of this paper.

\

\noindent $\bullet$ {\bf $\alpha$-attractor models}\\
The $\alpha$-attractor model \cite{Ferrara:2013rsa,Kallosh:2013yoa,Galante:2014ifa,Kallosh:2013hoa} is another benchmark model. Its inflation predictions are universal as long as the inflaton potential is smooth around the pole $\phi=\sqrt{6}\Mpl$ of the kinetic term.    
The predictions are 
\aln{
r\,\simeq\, \frac{12\alpha}{N^2}~,\quad n_s^{}
\,\simeq\, 1- \frac{9\alpha}{2N^2} - \frac{2}{N}~,\quad A_s^{}
\,\simeq\, \frac{V_0^{}N^2}{18\pi^2\alpha \Mpl^4}~, 
}
where $V_0^{}$ is the height of the inflaton potential at $\phi=\sqrt{6}\Mpl$. 
In the case of quartic potential Eq.~(\ref{inflaton potential}), we have 
\aln{
A_s^{} \=\frac{\lambda_\phi^{}(\Mpl)N^2}{2\pi^2\alpha} \qquad \therefore\ \frac{\lambda_\phi^{}(\Mpl)}{\alpha} \,\simeq\, 1.2\times 10^{-11}\times \left(\frac{60}{N}\right)^2~, \label{eq:alpha}
}
where the running of $\lambda_\phi^{}(\phi)$ is neglected.  
Thus, as in the non-minimal coupling case, $\alpha$ is determined as a function of $\lambda_\phi^{}(\Mpl)$. From $n_s$ data in \Eq{values of cosmological observables}, $\alpha \lesssim 10$ at 2$\sigma$ level, hence $\lambda_\phi(\Mpl) \lesssim 10^{-10}$. 
Thus, the allowed values of $\lambda_\phi(\Mpl)$ are somewhat smaller than those of the non-minimal coupling case.

As for the preheating, it was found that the effective mass of $\phi$ becomes tachyonic after inflation~\cite{Ema:2021xhq}, so that careful analyses are often necessary to estimate particle productions and resulting reheat temperature; see for example \cite{Krajewski:2018moi,Iarygina:2018kee,Iarygina:2020dwe} and  references therein.

\section{Probing small CMB amplitude by low-energy physics}  \label{sec:results}

\subsection{The connection between low and high energy physics}

\begin{figure}
\centering
\includegraphics[width=0.49\linewidth]{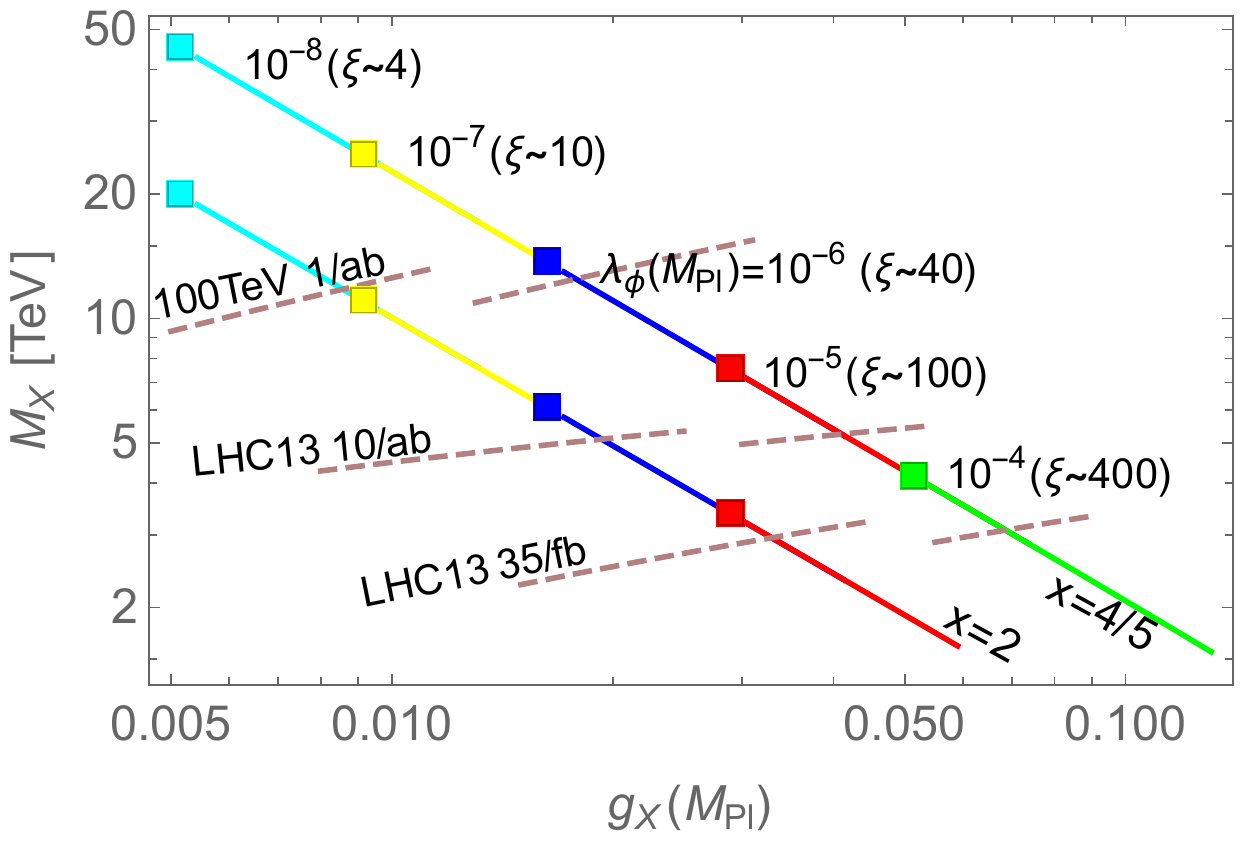}
\includegraphics[width=0.49\linewidth]{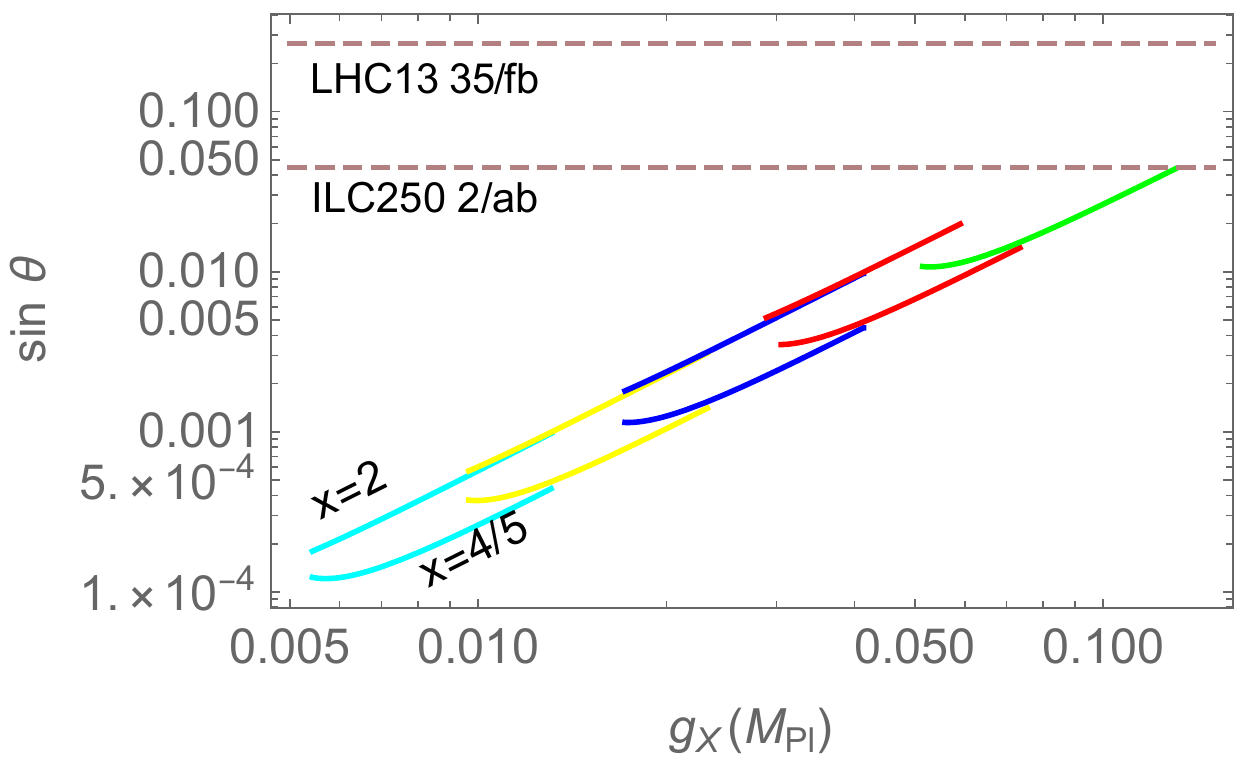}
\caption{ \label{fig:solparam} 
$M_X$ (left) and Higgs mixing $\sin \theta$ (right) predicted by solutions of correct EWSB, parameterized by $\lambda_\phi(\Mpl)$ and $g_X(\Mpl)$ for two different choices of $x=4/5$ and 2. For each $\lambda_\phi$ shown as different color (also marked with $\xi$ value for correct CMB amplitude), the minimum $g_X$ and corresponding maximum $M_X$ (and minimum $\sin \theta$) are marked as square dots; large-$g_X$ regions of different colors overlap. Recasted collider bounds from $Z^\prime$ searches and precision Higgs coupling measurements are shown as dashed lines. $\lambda_{h\phi}(\Mpl)=0$.
}
\end{figure}

The most crucial reason for the existence of this intriguing connection is that for given $\lambda_\phi(\Mpl) \ne 0$ there exists a \emph{maximum} $M_X$ (strongest low-energy constraints among many observables) consistent with successful EWSB. Consequently, a definite range of $\lambda_\phi(\Mpl)$, which is directly related to the CMB observables at the inflation scale, can be probed with low-energy constraints on $M_X$.

The explanation begins with the existence of the minimum $g_X(\Mpl)$ for given $\lambda_\phi(\Mpl) \ne 0$. A smaller $g_X(\Mpl)$ generally induces smaller (negative) RG corrections to $\lambda_{h\phi}$ ($\lambda_{h\phi}(\Mpl) =0$) so that a larger $v_\phi$ is needed to produce a correct $v_{\rm EW} = 246$ GeV from \Eqs{eq:mh2}{eq:vew}
\beq
v_{EW} \,\sim\, \sqrt{-\frac{\lambda_{h\phi}(v_\phi)}{\lambda_h(v_{EW})}} \, v_\phi \,\lesssim \, v_\phi,
\label{eq:EWfromportal} \eeq
where the last inequality is due to $|\lambda_{h\phi}| \ll \lambda_h \sim 0.1$. 
A larger $v_\phi$ needs larger $24 g_X^4(\Mpl) - 4y_N^4(\Mpl)$ in the dimensional transmutation of $v_\phi$ in \Eq{eq:vphi}.
In any case, too small $g_X(\Mpl)$ would require too large $v_\phi \gtrsim \Mpl$ to be realized in this model, hence the existence of the min $g_X(\Mpl)$\footnote{Note that \Eq{eq:vphi} depends directly on $\lambda_\phi(\Mpl)$ so that this argument does not directly apply to flatland scenarios with $\lambda_\phi(\Mpl)=0$.}.
The existence is proved numerically in \Fig{fig:solparam}, where the min $g_X(\Mpl)$ is marked with a square dot for each $\lambda_\phi(\Mpl)$ and for each $x=4/5$ and $2$. 
The values of the min $g_X(\Mpl)$ depend weakly on $x$ as the exponential dimensional transmutation of $v_\phi$ does not strongly depend on $x$.

The min $g_X(\Mpl)$ leads to the max $M_X$ for given $\lambda_\phi(\Mpl)$. Is it because the smaller $g_X(\Mpl)$ corresponds to the larger $v_\phi$, as discussed (see also \cite{Chun:2013soa}). Since $v_\phi$ depends exponentially on $g_X$, the resulting $M_X = 2 g_X(v_\phi) v_\phi$ is larger (Higgs mixing smaller). Thus, the min $g_X(\Mpl)$ is translated to the max $M_X$, as also proved numerically in \Fig{fig:solparam}. Moreover, the larger $v_\phi \gg v_{\rm EW}$ leads to smaller Higgs mixing corrections to Higgs couplings.
The values of max $M_X$ depend on $\lambda_\phi(\Mpl)$ and $x$. First, a larger $\lambda_\phi(\Mpl)$ typically needs a larger $g_X(\Mpl)$ to RG-drive $\lambda_\phi$ negative (for symmetry breaking). This also necessarily induces a larger portal mixing $|\lambda_{h\phi}(v_\phi)|$, thus a smaller $v_\phi$ yields a correct EW scale via \Eq{eq:EWfromportal}. Thus, larger $\lambda_\phi(\Mpl)$ predicts smaller max $M_X$ (and larger Higgs mixing).
Second, the $x$ dependence arises mainly from the running of $\lambda_{h\phi}$
\beq
(4\pi)^2 \beta_{\lambda_{h\phi}} \,\simeq\, 12 g_X^4 x^2.
\label{eq:betaportal} \eeq
Roughly speaking, the larger $x$, the larger $|\lambda_{h\phi}(v_\phi)|$ at $v_\phi$, so that a smaller $v_\phi$ can induce a correct $v_{EW}$, resulting in smaller max $M_X$. These behaviors are as shown in \Fig{fig:solparam}.

This is the main feature of the models where inflation and CW mechanism of EWSB are induced by a common particle.

\subsection{Main results}

It turns out that the constraints on $M_X$ (from $Z^\prime$ collider searches) provide the strongest probe. The current bound on the mass of $Z^\prime$ having the same interactions as the SM $Z$ is 4.5 TeV from LHC13 35/fb~\cite{Sirunyan:2018exx,Aaboud:2017buh}. It is recasted to the bound on $M_X$ for the same LHC13 35/fb, high-luminosity upgrade of LHC13 with 10/ab, and future 100 TeV $pp$ collider with 1/ab; these are shown in \Fig{fig:solparam} as dashed lines for each $x$. The recasted mass bound is where the same number of signal events are produced as the current bound; it is based on the assumption that signal and background compositions and cut efficiencies remain similar, which is reasonable for narrow resonance searches. The production ($q\bar{q}$-initiated) times BR ($e^+e^-, \mu^+ \mu^-$) is rescaled according to $x$-dependent charges, and the parton luminosity is taken from MSTW2008NLO~\cite{Martin:2009iq}.

$Z^\prime$ searches provide a meaningful probe of the inflation sector. For the model with $x=2$ shown in \Fig{fig:solparam}, a 100 TeV collider can probe $\lambda_\phi(\Mpl) \gtrsim 10^{-7}$ ($\xi \gtrsim10$) \emph{definitely}. This means first that larger values of $\lambda_\phi$ and $\xi$ cannot induce correct EWSB while being consistent with CMB and $Z^\prime$ searches. But this does not mean that a whole parameter space with a smaller $\lambda_\phi(\Mpl)$ can explain all CMB, EWSB and $Z^\prime$; rather, there exists some working parameter space, which usually yields small $g_X$ and heavy $M_X$. 
In this sense, we will say that a 100 TeV collider with 1/ab has a (definite) reach down to $\lambda_\phi(\Mpl) \simeq 10^{-7}$ and $\xi \sim 10$ for $x=2$; less natural size of $\xi \gtrsim 10$ will be more strongly constrained by 100 TeV collider searches.

\begin{figure}
\centering
\includegraphics[width=0.62\linewidth]{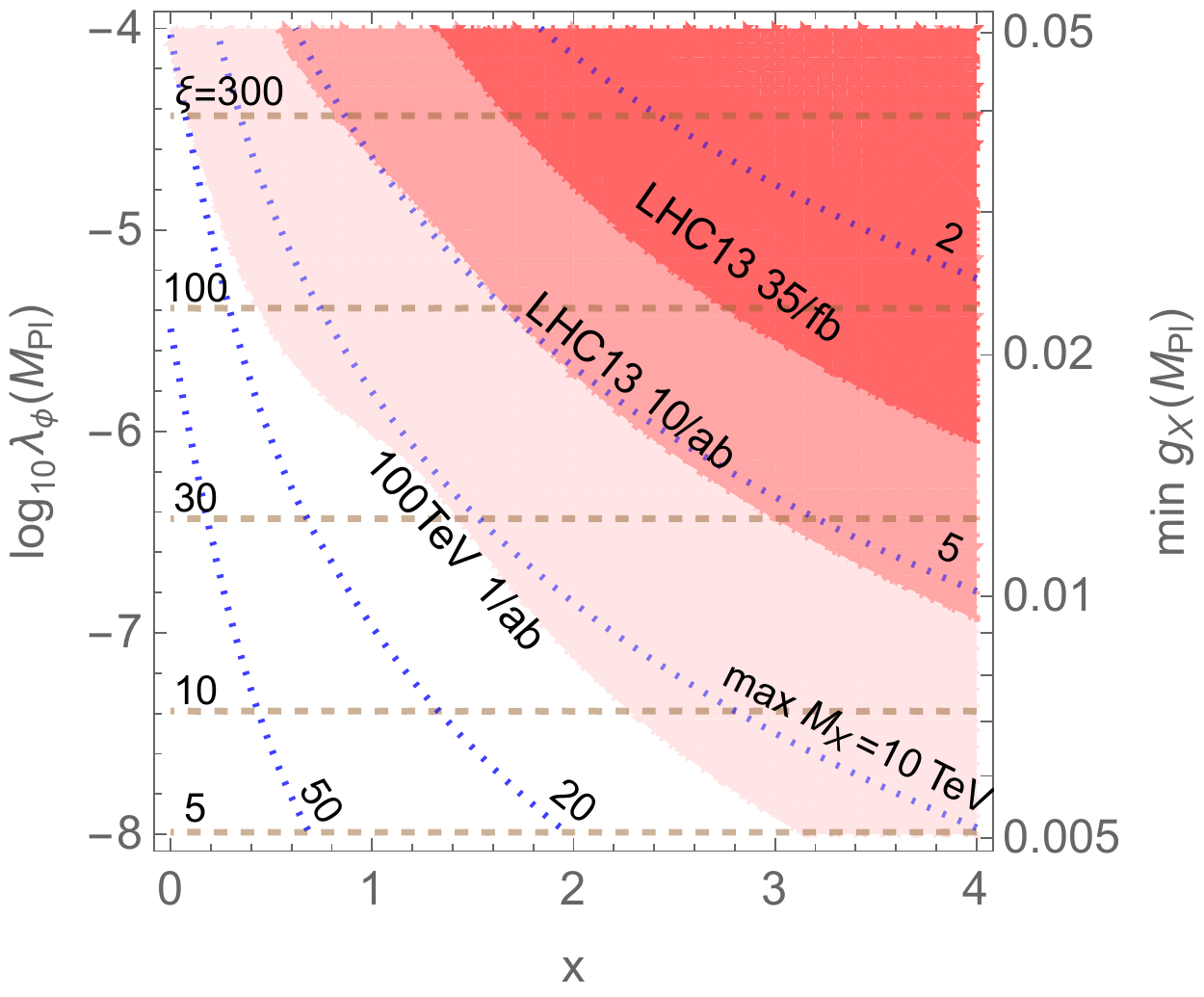}
\caption{ \label{fig:const} 
Collider constraints on the parameter space of inflation with non-minimal coupling $\xi$ and Coleman-Weinberg Higgs mechanism. On each point on the plane of $\lambda_\phi (\Mpl)$ and $x$, minimum $g_X(\Mpl)$ for correct EWSB and corresponding maximum $M_X$ (blue dashed) are used to obtain collider constraints from $Z^\prime$ searches. Red shaded regions are definite bounds, within which no parameter space can induce correct EWSB while being consistent with $Z^\prime$ searches. From darkest to lightest shaded regions are the bounds from current LHC13 35/fb, high-luminosity LHC13 10/ab projection, and future 100 TeV $pp$ collider 1/ab projection. Also shown as horizontal dashed lines are the required value of $\xi$ for correct CMB amplitude as in \Eq{eq:xi}; the most natural range of $\xi \lesssim {\cal O}(1-100)$ is preferred. $\lambda_{h\phi}(\Mpl)=0$. 
}
\end{figure}

\Fig{fig:const} shows such \emph{definite} bounds, more generally in the plane of $x-\lambda_\phi(\Mpl)$. Shaded regions are such excluded regions by recasted bounds on $M_X$; for given $\lambda_\phi(\Mpl)$ and $x$ within these regions, the max $M_X$ is lighter than the recasted bounds. Note that both recasted bounds and max $M_X$ (dotted contours) vary with $x$ and $\lambda_\phi(\Mpl)$ as well as with collider spec. Also overlapped are the contours of $\xi$ (horizontal dashed) for the correct CMB amplitude. In general, $\xi$ can be more strongly constrained for larger $x$, as discussed; a pure $B-L$ model with $x=0$ is much harder to probe. Due to this $x$-dependence, it is not appropriate to find a strict upper bound on $\xi$.
But we conclude that a large part of $\lambda_\phi \gtrsim {\cal O}(10^{-4} - 10^{-6})$ or $\xi \gtrsim {\cal O}(10-100)$ can be probed with current LHC13 35/fb and LHC13 10/ab; and $\xi \gtrsim {\cal O}(1-10)$ with 100 TeV $pp$ collider 1/ab. As discussed in \Sec{sec:inflationmodel}, the allowed range of $\xi$ may be considered most natural. Likewise, in the $\alpha$ attractor model, using \Eqs{eq:xi}{eq:alpha}, we conclude that a large part of $\alpha \gtrsim 10^6-  10^5$ and $10^5 -10^3$ can be probed, respectively.

\medskip
Lastly, Higgs-related physics gives weaker bounds but can still be relevant.  Higgs physics is modified by small portal mixing $\lambda_{h\phi}(v_\phi) = 10^{-9} \sim 10^{-4}$, for $g_X = 0.005 - 0.1$ respectively. The resulting Higgs mixing angle is
\beq
\sin \theta \, \simeq\, \lambda_{h\phi}(v_\phi) v_\phi v_{EW}/m_h^2 \, \sim \, \sqrt{\lambda_{h\phi}(v_\phi) /2\lambda_h} \= 10^{-4} \sim 10^{-1}
\eeq 
as also shown in \Fig{fig:solparam} right panel. But these are too small to be probed with current LHC precision $\sin \theta \lesssim 0.27$ from ATLAS+CMS with 35/fb~\cite{Aad:2019mbh,Sirunyan:2018koj} as well as with expected precision $\sin \theta \lesssim 0.045$ from ILC 250 S2 stage with 2/ab~\cite{Barklow:2017suo,Bambade:2019fyw}. 
In addition, $\phi$ is expected to be light $M_\phi \simeq \sqrt{\frac{6}{11}\lambda_\phi(v_\phi)} v_\phi = {\cal O}(0.1-10)$ GeV for most parameter space, but $h\to \phi \phi$ decay rate is still too small
$\Gamma(h \to \phi \phi) \, \simeq\, \frac{\lambda_{h\phi}(v_\phi)^2 v_{EW}^2}{32\pi m_h} \sqrt{1-\frac{4m_\phi^2}{m_h^2}} \,\lesssim \, 10^{-3} \,{\rm MeV}$ (or, its branching ratio $\lesssim 10^{-3}$)
to be probed even at ILC 250, whose expected precision on invisible decay branching ratio is $\sim 0.003$~\cite{Barklow:2017suo,Bambade:2019fyw}.
LHCb, BaBar, and Belle are sensitive to GeV-scale dark photons with its interaction strength $\epsilon \gtrsim 10^{-3}- 10^{-4}$~\cite{Aaij:2019bvg}, but the $\phi$-lepton interaction yields too small $\epsilon = \frac{y_e}{e} \sin \theta \simeq 9.2 \times 10^{-6} \sin \theta \ll 10^{-3}$ in this scenario. Nevertheless, as discussed, Higgs-related physics can be important for proper reheating.

\subsection{Variation with non-zero portal mixing}

\begin{figure}
\centering
\includegraphics[width=0.50\linewidth]{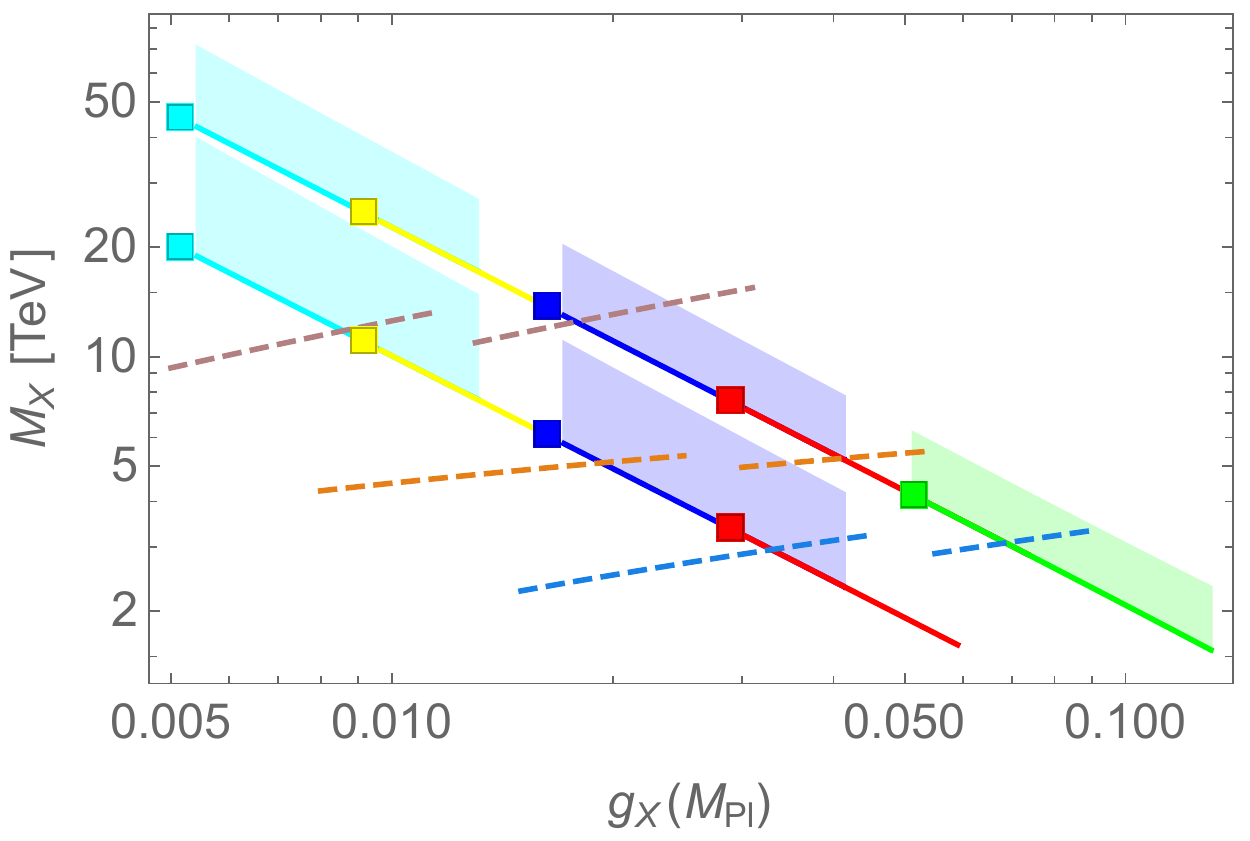}
\includegraphics[width=0.49\linewidth]{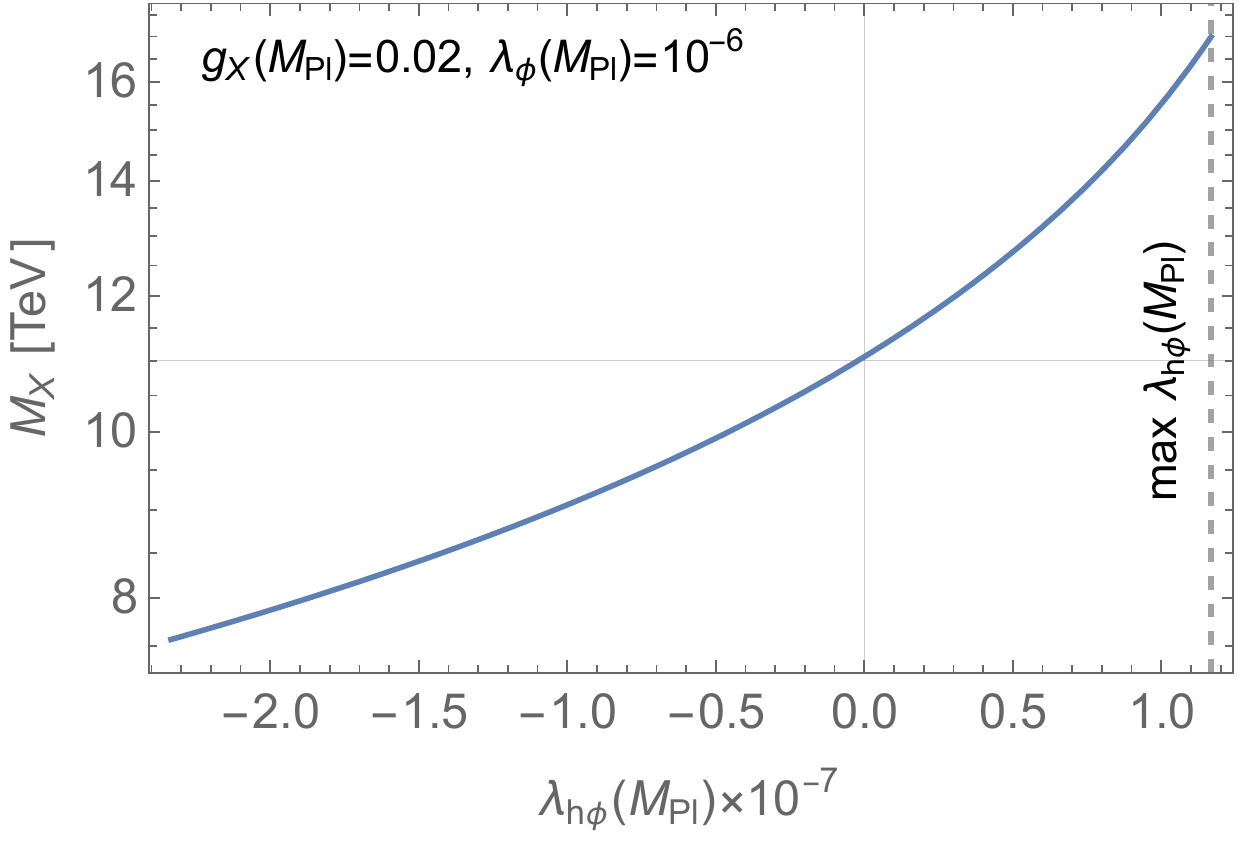}
\caption{ \label{fig:variportal} 
{\bf (Left)}: Same as \Fig{fig:solparam} but showing variations of low-energy predictions with $\lambda_{h\phi}(\Mpl) > 0$ (shaded regions). The shaded regions still have definite ranges determined by maximum positive $\lambda_{h\phi}(\Mpl)$ in \Eq{eq:maxportal} that can induce correct EWSB, which yields somewhat heavier $M_X$ for given $g_X(\Mpl)$.  {\bf (Right)}: An example $M_X$ prediction as a function of $\lambda_{h\phi}(\Mpl)$ for fixed $g_X(\Mpl)=0.02$ and $\lambda_\phi (\Mpl)=10^{-6}$. The maximum $\lambda_{h\phi}(\Mpl)$ consistent with EWSB is marked as a vertical dashed line. See text for more discussion.
}
\end{figure}

We now assess the variation with $\lambda_{h\phi}(\Mpl)\ne 0$. It cannot be arbitrarily large positive for given $g_X(\Mpl)$. Since $\lambda_{h\phi}(v_\phi) <0$ has to be negative to induce EWSB, its RG running should be large enough
\beq
\Delta \lambda_{h\phi} \,\simeq\, \beta_{\lambda_{h\phi}} \cdot \log \left( \frac{v_\phi}{\Mpl} \right) \, \simeq \, \frac{12 g_X^4 x^2}{(4\pi)^2} \cdot {\cal O}(10) \, \gtrsim \, \lambda_{h\phi}(\Mpl).
\label{eq:maxportal} \eeq
Thus, for given $g_X(\Mpl)$ (and $\lambda_\phi(\Mpl)$), there exists a maximum $\lambda_{h\phi}(\Mpl)$ inducing correct EWSB. Importantly, this still yields a max $M_X$ for the given parameters because larger positive $\lambda_{h\phi}(\Mpl)$ yields a smaller $|\lambda_{h\phi}(v_\phi)|$, hence requiring a larger $v_\phi$, yielding heavier $M_X$. The max $M_X$ will be larger for $\lambda_{h\phi}(\Mpl)>0$, but the maximum still exists. In the other limit of negative $\lambda_{h\phi}(\Mpl)<0$, the max $M_X$ is smaller so that more strongly constrained. These are numerically proved in \Fig{fig:variportal}; the modified max $M_X$ is shown in the left panel as shaded regions, and the max $M_X$ as a function of $\lambda_{h\phi}(\Mpl)$ is shown in the right panel where the max positive $\lambda_{h\phi}(\Mpl)$ for EWSB is marked as a vertical line. The new max $M_X$ is larger by a factor 1.5 -- 2.0.

\begin{figure}
\centering
\includegraphics[width=0.47\linewidth]{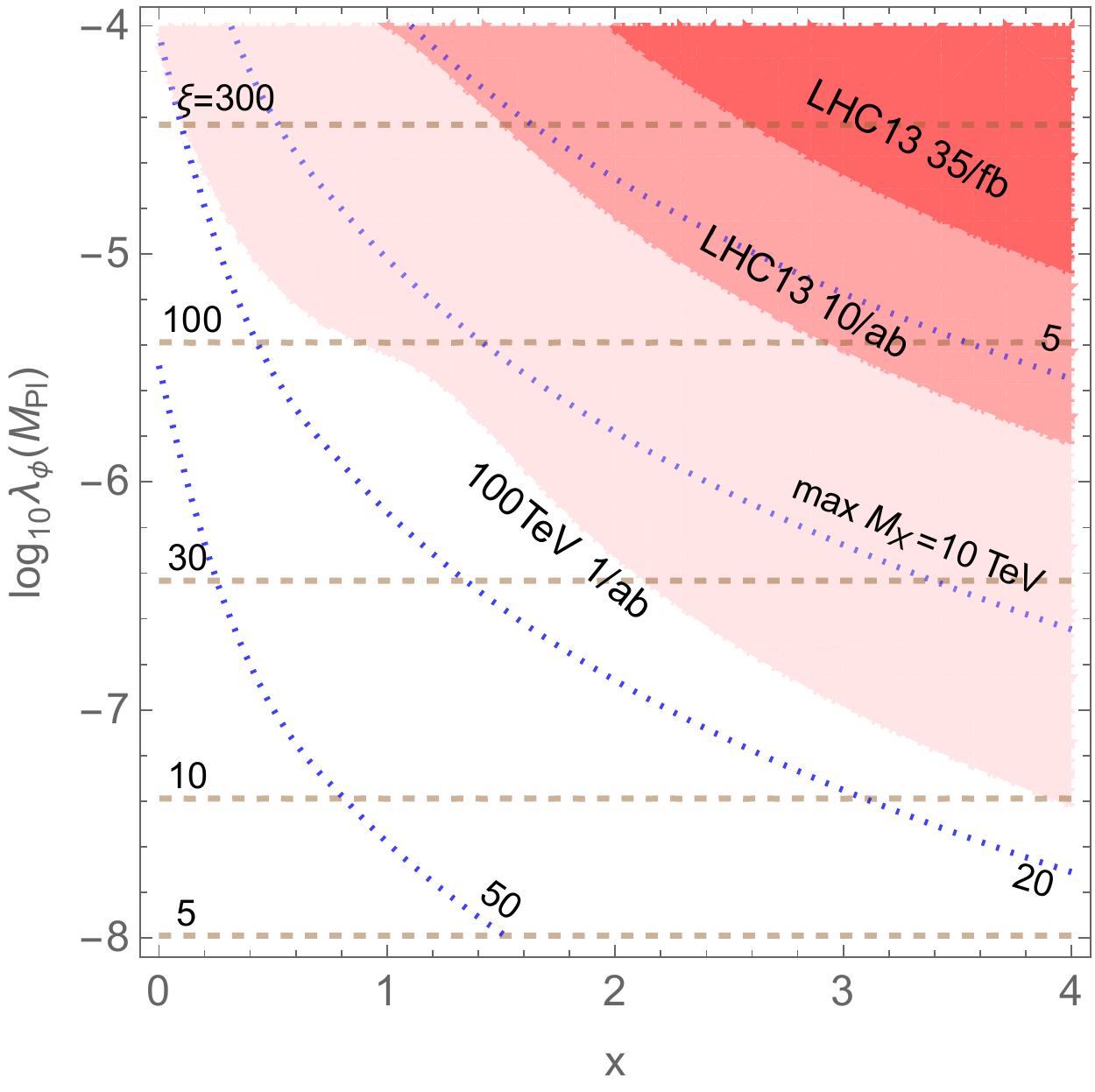}
\includegraphics[width=0.47\linewidth]{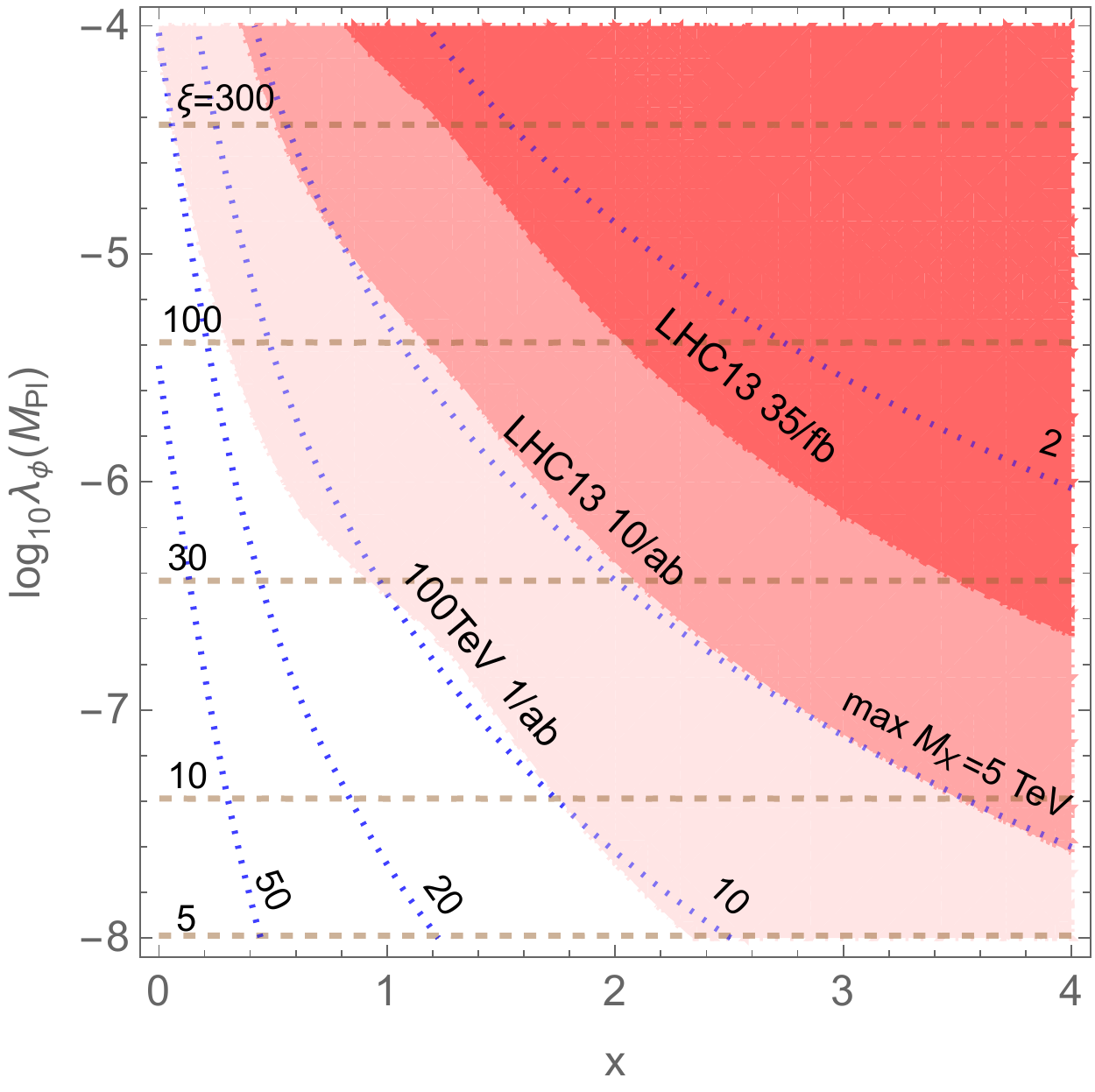}
\caption{ \label{fig:const-pm} 
Same as \Fig{fig:const} but now with $\lambda_{h\phi}(\Mpl) \ne 0$: maximum positive $\lambda_{h\phi}(\Mpl)>0$ for EWSB as in \Eq{eq:maxportal} and \Fig{fig:variportal} yielding heavier max $M_X$ (left panel), and negative $\lambda_{h\phi}(\Mpl)<0$ with the twice magnitude yielding lighter max $M_X$ (right panel). 
}
\end{figure}

In \Fig{fig:const-pm}, we show definite bounds with max positive $\lambda_{h\phi}(\Mpl)$ (worst constraints) as well as with negative $\lambda_{h\phi}(\Mpl)$ with the twice magnitude (stronger constraints). Again, shaded regions are definite bounds, within which no parameter space can be consistent with all CMB, EWSB and collider searches. For the former case, current LHC13 35/fb can still probe a large part of $\xi \gtrsim {\cal O}(100)$, and future high-luminosity LHC13 and 100 TeV 1/ab can probe a large part of $\xi \gtrsim 100-10$. 
But this is for the max $\lambda_{h\phi}(\Mpl)$ yielding the worst constraints; but $\lambda_{h\phi}$ would more likely take a smaller or negative value. In such a case (right panel), the constraints/prospects are stronger, and the majority of $\xi \gtrsim 1-10$ can be probed with future 100 TeV 1/ab, and current LHC13 35/fb can already probe a large part of $\xi \gtrsim {\cal O}(100)$. In any case, smaller $\xi$ will be preferred by collider experiments.

\section{Summary}

We have shown that the parameter space for the small CMB amplitude can be probed by collider experiments. In the inflation model with non-minimal coupling $\xi$, the absence of collider signals of $Z^\prime$ at LHC 13 and 100 TeV $pp$ will prefer a smaller and more natural size of $\xi \lesssim 1-100$ with $\lambda_\phi(\Mpl) \lesssim 10^{-4} - 10^{-8}$ even though $\xi$ could equally well take a much larger and less natural value to explain the CMB amplitude. Another possibility of $\lambda_\phi(\Mpl) \sim {\cal O}(1)$ with $\xi \sim 10^5 \gg1$ will be strongly constrained. A similar conclusion for an $\alpha$ attractor model is obtained, albeit more weakly.

This probe was possible in models where an inflaton $\phi$ also induces EWSB via perturbative quantum corrections (CW mechanism). The restrictions on the $\phi$ potential from well-measured high-energy inflation and low-energy electroweak physics lead to intimate correlations, in particular between low-energy $M_X$ (collider observables) and high-energy $\lambda_\phi(\Mpl)$ (CMB amplitude $A_s \propto \lambda_\phi(\Mpl)$).  More crucially, there exists an upper range of $M_X$ prediction for each value of $\lambda_\phi(\Mpl)$, so that definite collider constraints on $\lambda_\phi(\Mpl)$ could be derived.

This work not only proves an interesting possibility of probing the inflation sector with low-energy experiments, but may also relate the absence of new physics signals to the physics at a disparate energy scale. Although this connection does not explain why our universe is realized with such small CMB amplitude, the fact that such particular realization at high-energy scale also has consequences in low-energy physics is intriguing enough, worthwhile to be pursued in a more varieties of models.

\begin{acknowledgments}
We are supported by Grant Korea NRF-2019R1C1C1010050, 2019R1A6A1A10073437, 
and SJ also by POSCO Science Fellowship. 
\end{acknowledgments}

\bibliography{Bibliography}
\bibliographystyle{utphys}

\end{document}